# Robust Technique for Representative Volume Element Identification in Noisy Microtomography Images of Porous Materials Based on Pores Morphology and Their Spatial Distribution

Grigoriev M.[1], Khafizov A.[2,3], Kokhan V.[4], Asadchikov V.[3]

[1]Institute of Microelectronics Technology and High-Purity Materials RAS, Chernogolovka, Russia
[2]National Research Nuclear University MEPhI, Moscow, Russia
[3]FSRC "Crystallography and Photonics" RAS, Moscow, Russia
[4]Smart Engines Services LLC, Moscow, Russia

## ABSTRACT

Microtomography is a powerful method of materials investigation. It enables to obtain physical properties of porous media non-destructively that is useful in studies. One of the application ways is a calculation of porosity, pore sizes, surface area, and other parameters of metal-ceramic (cermet) membranes which are widely spread in the filtration industry. The microtomography approach is efficient because all of those parameters are calculated simultaneously in contrast to the conventional techniques. Nevertheless, the calculations on Micro-CT reconstructed images appear to be time-consuming, consequently representative volume element should be chosen to speed them up. This research sheds light on representative elementary volume identification without consideration of any physical parameters such as porosity, etc. Thus, the volume element could be found even in noised and grayscale images. The proposed method is flexible and does not overestimate the volume size in the case of anisotropic samples. The obtained volume element could be used for computations of the domain's physical characteristics if the image is filtered and binarized, or for selections of optimal filtering parameters for denoising procedure.

## 1. INTRODUCTION

Microtomography (Micro-CT) method is one of the most powerful approaches to investigate the properties of porous media these days. There are a lot of scientific and industrial realms, where Micro-CT is almost irreplaceable, e.g. the filtration industry, medicine, materials science, etc. In particular, in the filtration industry they utilize polymeric, ceramic, metallic, and metal-ceramic (cermet) membranes that represent a porous material with micro-, ultra- and nano-scale pore throats, which enables to separate molecules of different size and characteristics [1]. Apart from chemical composition, the efficiency of purification depends on such parameters as pore size distribution, effective porosity and the specific surface area of pores. All mentioned parameters could be obtained via conventional techniques, e.g. mercury porosimetry, but calculation of them will be performed in one go using Micro-CT binarized reconstructed image [2], which is much more convenient. Moreover, due to the tomography approach is not a destructive method, samples could be deployed further for other experiments.

However, the main drawback of Micro-CT is the low speed of calculations with reconstructed volume that is caused by the big size of experimental data [3].

Representative elementary volume (REV) is the smallest homogeneous part in heterogeneous domain, from which properties or parameters of materials become independent of size. There are two types of it – deterministic and statistical that are correspondingly based on physical and statistical terms [4]. For Micro-CT measurements of chaotic media [5], statistical theory is more universal and applicable. It implies work with averaged parameters throughout the whole sample domain to find REV shape, then REV could be truncated from any part of the sample.

However, due to uncertainty of REV position the overestimation of the size may occur. If the position is chosen incorrectly, the larger REV will be required to capture all necessary features. Hence, the usage of the representative volume element (RVE) instead is more practical. It is a basic concept in mechanics where it is used for estimating average properties of heterogeneous solid [6,7]. Although notions are close, RVE has determined position whereas REV concerns only the size of the representative element. Eventually, in this paper, under REV we mean the shape of RVE hereafter.

In the current paper we purpose the concept of RVE applied to the field of Micro-CT. It could be useful for investigation of the morphology of porous materials, its physical properties, but also could be used for image denoising (investigation of optimal filtering parameters [8] via the solution of an optimization task). The paper emphasizes RVE finding in rough, non-filtered Micro-CT images because most of the already known methods, e.g. purposed in [4,9,10], will not manage to find REV or RVE in noised samples.

## 2. MATERIALS AND METHODS

### 2.1. Phantoms

To examine the proposed techniques, phantoms (artificial samples) were used. They represent ordered numerical dataset consisting of "0" and "1" values that matches to pore and material respectively.

In this paper, PoreSpy python library was utilized as a generator [11] for some of the mentioned phantoms. It works with several parameters: the shape of a phantom (size of a rectangular image in voxels), porosity (ratio between total pore-voxel number and phantom volume in voxels), and blobiness (controls the morphology of pores: the lower value this parameter has, the larger pore diameters are).

In our work [4] it was described how we choose all of those PoreSpy parameters in detail. According to our previous studies, the essential width and length of the samples are $1000 \times 1000$ voxels, porosity should be 0.1 or 0.4, blobiness – 2 or 5. In this paper only one phantom with large pores is enough to demonstrate the work of the algorithm. Hence only sample with porosity 0.4 and blobiness 2 is considered (Fig 1a). It should be considered that in contrast to the work [4], here black color denotes pore volume and white – material. We chose the height of the phantom along the rotation axis – 300 voxels, i.e. the shape is $1000 \times 1000 \times 300$ voxels overall.

To make our tests more explicit, ideally homogeneous porous specimens were simulated using PoreSpy tools. It mimics the primitive crystal structure where void space of the structure was replaced by the material; and on the contrary, molecules typify the void pore space. The phantom is demonstrated in Fig 1b.

In conclusion, we developed a new type of phantom where pores shape is anisotropic and could be explicitly set in any direction. This sample is important to show up the divergence of parallelepiped edge sizes of RVE in the case of anisotropic samples. The example is shown in Fig 1c.

Homogeneous and anisotropic specimens' shapes are only $1000 \times 1000 \times 1$ voxels to save computational time. According to [4], one pixel has a size of 1 μm.

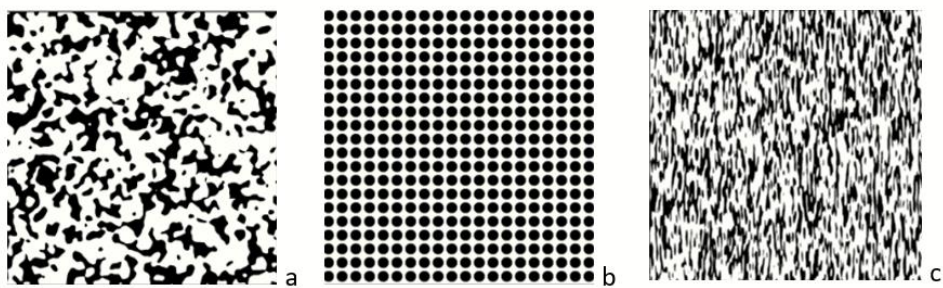

Figure 1. Phantoms' cross-sections: a – similar to real sample; b – highly periodic; c – anisotropic. Pores are black.

### 2.2. Simulation of a microtomography experiment

Microtomography experiment is an imaging technique that enables to non-destructively investigate the inner structure of an observed specimen. In the current paper, we consider that all the experiments are conducted in parallel X-ray beams only, which is described in work[12]. Here, a brief description of the key points is provided.

Firstly, the sinogram is calculated through the direct Radon transform of the samples. Mathematically it means, that voxels of materials were summed up in rotated, with fixed step, direction (direction of the X-ray beam). Hence, if that sinogram is multiplied by a physical size of a voxel $d$, e.g. in mm, the material physical length, passed by the beam, is obtained as a result. Consequently, if we imagine CMOS detector capturing passed X-ray photons, then, for nominal X-ray source intensity $I_0$ (it means number of photons per frame that is captured by one voxel of the detector), it will fix the following photon distribution corresponding to a sinogram voxel $s$:

$$I = Pois(I_0) \cdot exp(-\mu s d)$$

where $\mu$ – is attenuation coefficient; $Pois(I_0)$ is a generated Poisson random number (mean of the Poisson distribution is equal to $I_0$). Here, Poisson distribution emulates the experimental noise appearing in the reconstructed images; and the more is $I_0$ value the less is the noise effect.

Afterward, the photon discrete nature and scattering factor are taken into account, and intensities are smoothed via convolution with gaussian kernel $g(\sigma = 0.5)$ [13] and then rounded to integers:

$$I_{real} = \lfloor I * g(\sigma) \rfloor$$

Finally, the derived real intensity distribution is turned back to the sinogram space:

$$s_{real} = -\frac{1}{\mu d} ln \frac{I_{real}}{Pois(I_0)}$$

The described procedure was performed for all initial sinogram voxels. Next, inverse Radon transform [12,14] of $s_{real}$ led to the noised Micro-CT reconstructed image that is close to the real experimental data.

For our numerical experiments $Al_2O_3$ was chosen as the material for our phantoms because of the high popularity of this material in cermet production [15]. Assume that anode of utilized X-ray source is molybdenum with mean photon energy equal to 33 keV, consequently the attenuation coefficient in this case is 0.25 mm$^{-1}$ [15]. From section 2.1, a voxel size is $10^{-3}$ mm. In order to save computation time, only 180 Radon projections in each sample were captured. Examples of how the proposed simulation works with samples that are similar to one demonstrated in Table 1.

Table 1. Tomography reconstructed images of the samples (porosity – 0., blobiness 2) for diverse X-ray source intensity $I_0$

| $I_0$ | 100 | 1000 | 3000 |
|---|---|---|---|
| Micro-CT reconstructed images (cross-section). Shape: $1000 \times 1000 \times 300$ voxels | 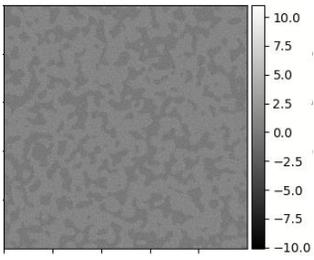 | 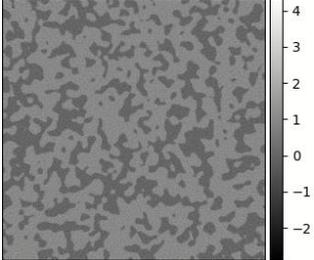 | 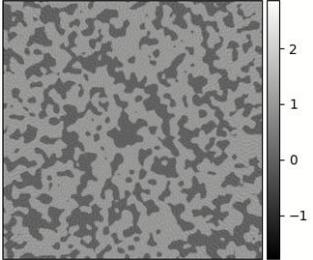 |
| Central part after unbalanced Otsu binarization [16] of the whole volume (cross-section). Shape: $100 \times 100 \times 100$ voxels | 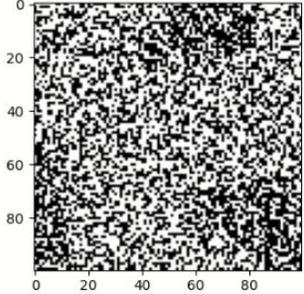 | 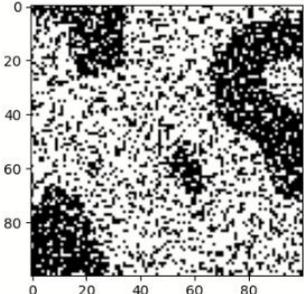 | 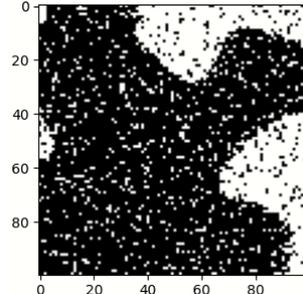 |

## 2.3. REV determination algorithm

Initially, we are working with 3d noised gray Micro-CT reconstructed images. We cannot operate any physical parameter such as porosity, pore size distribution, and others because we did not label each voxel as pore or material yet. Even if images are binarized somehow, the problem will not be solved in cases when the noise factor is high. To overcome this issue, we propose the following strategy.

Assume, that we have 3d Micro-CT images which represents a set of voxels $P$ in the shape of a rectangular parallelepiped. Our goal is to find the RVE $p \subset P$. Two special cases should be taken into consideration: if the reconstructed image represents a constant value in all voxels the REV should be 1 voxel; the minimum REV should be seen if voxel intensities are completely random (white noise) because there even enlarged volume will not be informative. The less the noise effect is, the closer RVE size should be to the average size of main features – pores or materials.

Firstly, the task should be simplified: instead of seeking REV in 3-dimensional space, let's try to do it in 1 dimensional one. On this purpose, we randomly choose some rows of voxels along any cartesian direction. Example of possible positions of rows in the reconstructed image depicted in Fig. 2a; the blue row section is visualized as a function dependence plot in Fig 2b.

As we see from Fig. 2b, the function is wavy because of tomography noise existence. To filter the noise with the lowest amplitude all sections should be binarized. Although this step could be omitted that will lead to slight unpredictable deviations of the finale REV values. After binarization, the function will look similar to a step one that is shown in Fig. 2c. It is clear to see that there are some sharp peaks and throughs, e.g. highlighted by dashed red rectangle, which are highly likely artifacts caused by the significant tomographic noise divergence. More explicitly it is seen when 3d reconstructed image is binarized, then "levitating stones" appear, i.e. a few voxels of material without any joint with the bulk or facets [17]. If the reconstructed image contains more than two phases, e.g. air, water, and solid, the binarization step could be replaced by the segmentation with multiple thresholds [18].

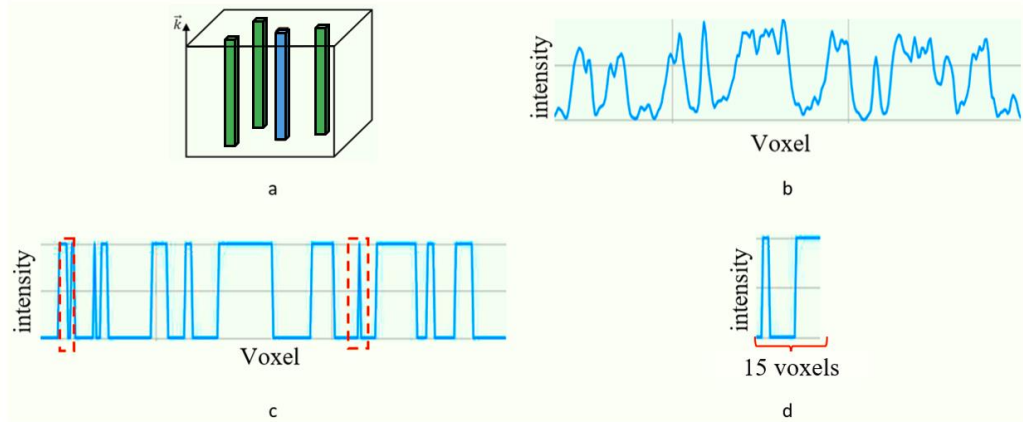

Figure 2. Demonstration of REV computation: a – voxel rows sampling; b – blue row intensity distribution in grayscale reconsructed image; c - blue row intensity distribution in binarized reconstructed image with noise artifacts (dashed red rectangles); d – row segment.

In fact, the key point in REV calculation is the search of the most repetitive pattern through the analysis of voxel rows which are parallel to a cartesian axis. The question is how it could be performed.

Let's denote repetitiveness $R$ of a row segment $L_1$ (e.g. illustrated in Fig. 2d), that belongs to a chosen set of voxel rows $M$, as the probability to coincide another row segment $L_2 \subset M$ in other position; and number of voxels in the segments $L_1$ and $L_2$ is equal (i.e. the have equal length). In other words, let's randomly truncate the rows in $N$ segments of equal length $L$ and find a number $n(\epsilon)$ of similar ones. It means that similarity $S(L_1, L_2) > \epsilon$, where $\epsilon$ is a precision parameter: $0 < \epsilon \leq 1$. Then:

$$R(\epsilon, L) = \frac{n(\epsilon, L)}{N}$$

We treat the similarity of the segments as:

$$S(L_1, L_2) = \begin{cases} |corr(L_1, L_2)|, if\ intensites\ of\ L_1\ AND\ L_2\ are\ not\ CONST \\ 1, if\ intensites\ of\ L_1\ AND\ L_2\ are\ CONST \\ 0, if\ intensites\ of\ L_1\ XOR\ L_2\ are\ CONST \end{cases},$$

where $corr(L_1, L_2)$ is the Pearson correlation coefficient [19]. It is taken the absolute value of the coefficient due to the invariance of numerical image representation for the filtering process: there is no difference whether pores assigned as 0 and stones as 1 or conversely. Thus, $S(L_1, L_2) = 0$ means that segments $L_1$ and $L_2$ are completely dissimilar, and 1 indicates the highest level of similarity. In non-periodic structure, such as our porous media, two segments could be considered as similar if the similarity is higher than a quite low constant value $0 < \epsilon \leq 1$.

In conclusion, varying $L$ from 2 voxels to the maximum number that one row contains, we compute dependency $R$ from $L$. In most cases, it is a monotonically decreasing function. In highly periodic samples it becomes parabolic or even degenerates to an almost constant value, but still, the minima are located far from the $L = 0$.

Next, we define the representative elementary size (RES) as the shortest length $L$ where the $R(\epsilon, L)$ minimum is being reached. It could be done by guiding the following two steps. Firstly, the plot is normalized from 0 to 1 by subtracting its minimum value and then dividing by the difference between the maximum and minimum. Secondly, by extending $L$ from 2 to its maximum, there should be found $R_{normed}(\epsilon, L)$ which is close to zero, possibly with a certain deviation $\xi$. Overall:

$$\text{RES} = \min(L), \text{ where } \frac{R(\epsilon, L) - \min(R(\epsilon, L))}{\max(R(\epsilon)) - \min(R(\epsilon))} < \xi$$

Thus, any length lower than RES is highly repetitive in relation to the remainder, i.e. RES contains all most repetitive segments. An example of RES and its possible size are illustrated in Fig. 2d. Now, the goal is to find the RES sizes for all cartesian directions specified in our reconstructed image in Fig. 2a, and then determine RVE as:

$$shape(RVE) = \left(RES_{\vec{x}}, RES_{\vec{y}}, RES_{\vec{z}}\right) = RES_{\vec{x}} \times RES_{\vec{y}} \times RES_{\vec{z}}$$

This strategy is sufficient because edge lengths are independent of each other that enables RVE to have not only a cubic shape, but also a rectangular with the highest edge length along anisotropy direction.

It should be also mentioned, that there is a weak point regarding the anisotropy directed along a image's diagonal. In that case, the RVE will be overestimated and will have a cubic shape. The only way to tackle this problem is the rotation of the image manually if it is needed (nonetheless, in noisy images RVE size will be small enough).

Utilization of the Pearson metric restricts the minimum size of the reconstructed image, it has to contain at least 4 voxels in one voxel row or multiple rows with not less than 2 voxels in the specified direction. If there is not enough number of voxels to make sampling in a chosen direction, we assume that RSE has the maximum possible length in that case. Therefore, when we deal with linear detector and reconstructed image has $1000 \times 1000 \times 1$ voxels, RVE could only have shape $x \times y \times 1$ voxels, where integers $x, y \leq 1000$.

Overall, the described algorithm is presented in the flowchart in Fig. 3 According to this algorithm, REV in such singular cases as pure noise, and constant function is $2 \times 2 \times 2$ voxels that confirms that theory is correct.

### 2.4. RVE positioning

In previous steps the shape determination strategy of RVE was described. In highly homogenous media, the found shape of representative elementary volume could be cropped from any place of the initial volume thanks to the high repetitiveness level. Contrary, in cases when the distribution of noise, pores, or other important features is not even enough, the strategy should be developed.

For example, let's consider a property $P$ as the most valuable in our sample, and it is unevenly distributed throughout the whole reconstructed volume $V$. Then, the most proper position of RVE could be selected in the place where the average of $P$ calculated inside the $V$ and RVE are the closest to each other:

$$|\langle P \rangle_V - \langle P \rangle_{RVE}| \to min$$

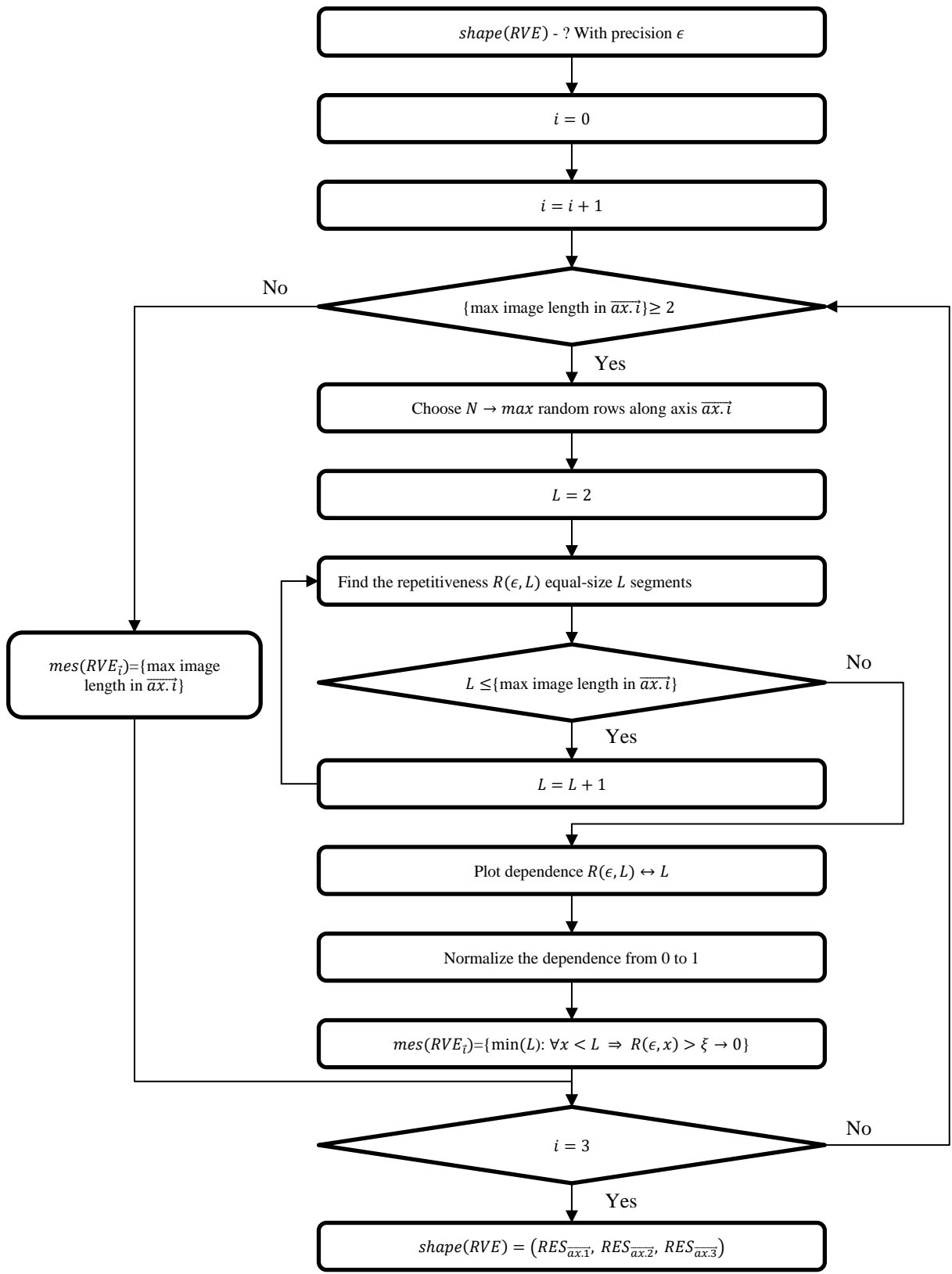

Figure 3. Determination of REV. Algorithm.

## 3. RESULTS AND DISCUSSIONS

In section 2.3 the algorithm about REV calculation is described. There we randomly take a fixed large number of rows along a direction and compare their segments using similarity criterion with precision $\epsilon$. In Fig. 4, it is illustrated how the RVE size changes depending on $\epsilon$ value and quantity of the voxel rows. All numerical experiments were conducted on the phantom shown in Fig. 1a (on the original image, without any tomography modeling in this step); calculations performed only in the horizontal axis and averaged, the vertical one was omitted.

As the plot suggests, the RVE length size is high even if $\epsilon = 0.9$ that means that our data does not contain a noise and has quite large pores.

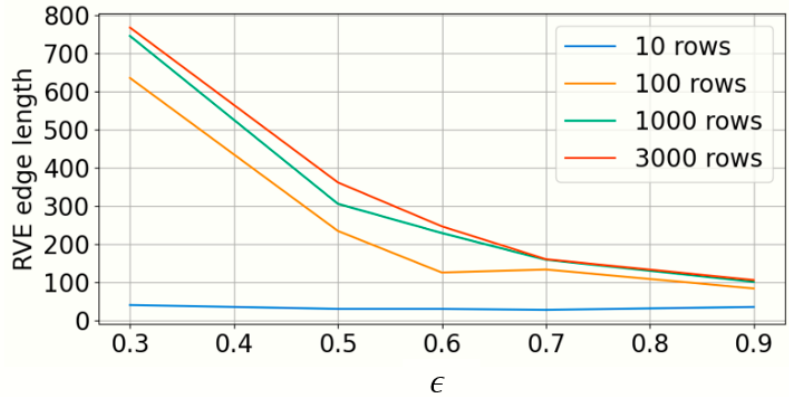

Figure 4. Dependency of REV from precision parameter $\epsilon$ and sampling frame.

After numerous launch of the script, lines assigned to 100, 1000, and 3000 rows are close to each other which means that sampling frame of 10 000 rows is comprehensive for our images (0.3% of total rows count that are accessible for sampling). The analysis is based on Fig. 5.

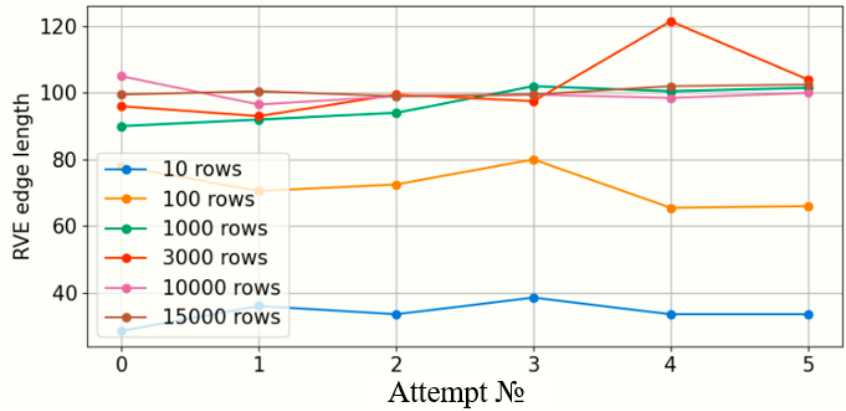

Figure 5. Stability of REV in relation to the number of selected rows for the phantom shown in Fig. 1a, $\epsilon = 0.9$.

Fig. 6 depicts how changes the repetitiveness in different samples described in section 2.1 when the length of the truncated segment varies from 2 voxels to the maximum value (1000 voxels). As it was predicted, the minimal function values are in a distance from $L = 0$. The parameter $\xi$, which is used to reduce the influence of calculation inaccuracy, is equal to 1%. The plot shows that RES is equal to 91 voxels in a chosen direction.

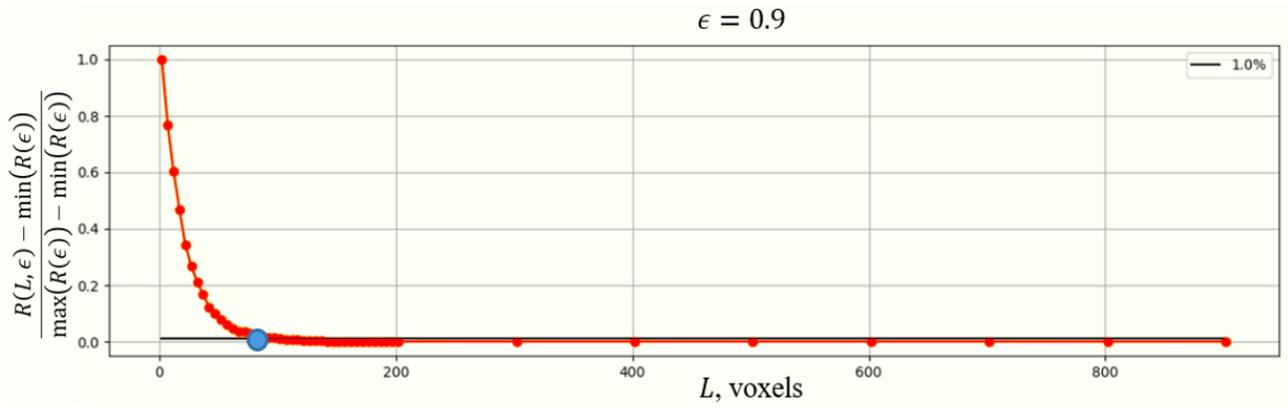

Figure 6. Repetitiveness depending on cropped segment size. The plot is normalized from 0 to 1, the blue dot corresponds to RES.

Table 2 contains REVs for generated phantoms and some tomographic reconstructed images outlined in section 2.2. It is clear to see that the most noised image has the lowest RVE size that completely fits the theory. Furthermore, in the case of the anisotropic sample, REV length in the anisotropy direction, along which the pores are elongated (Fig. 1c), is larger than in perpendicular directions, i.e. RVE is not overestimated. The RVE is presented in Fig. 7.

The table also tells that RVE shape decreases when the image becomes noisier; this fact also completely fits the theory.

Table 2. RVE shapes for phantoms and Micro-CT reconstructed images. Precision $\epsilon = 0.9$, 10 000 rows.

| Sample | | Reference | Phantom shape | Anisotropy | RVE shape |
|---|---|---|---|---|---|
| Similar to real phantom | | Fig 1a | $1000 \times 1000 \times 300$ | | $99 \times 91 \times 98$ |
| Micro-CT reconstructed images | $I_0 = 3000$ | Table 1 | $1000 \times 1000 \times 300$ | No | $21 \times 21 \times 22$ |
| | $I_0 = 1000$ | | | | $11 \times 11 \times 11$ |
| | $I_0 = 100$ | | | | $10 \times 10 \times 10$ |
| Anisotropic phantom | | Fig 1c | $1000 \times 1000 \times 1$ | Yes | $115 \times 28 \times 1$ |

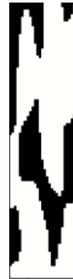

Figure 7. Representative volume element for the anisotropic phantom from Fig. 1c., $\epsilon = 0.9$

Figure 8 shows the correlation between the RVE size and precision parameter for noisy Micro-CT reconstructed image ($I_0 = 1000$). Appropriate $\epsilon$ parameter should be chosen considering the certain purpose of the RVE. For instance, if it is deployed to find optimal filtration parameters to get rid of levitating stones [17], the parameter should be low to widen the REV to the size up until it has a lot of levitating stones; otherwise, that optimization problem may not be solved.

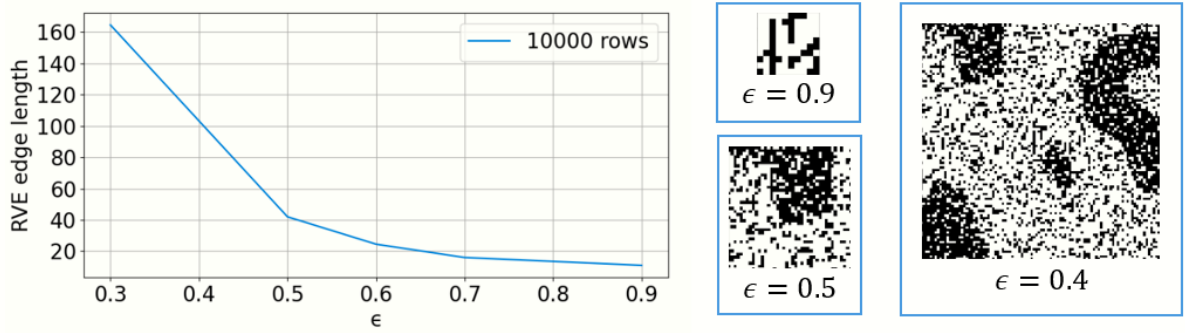

Figure 8. Correlation between REV and precision parameter $\epsilon$ in the Micro-CT reconstructed image ($I_0 = 1000$).

Finally, for the highly periodic sample from Fig. 1b the RVE, cropped from the random place is illustrated in Fig. 9. The crop center could be identified guiding the average porosity criterion and concept described in section 2.4 as well. It is also demonstrated in the figure. It is clear to see, that REV identified correctly and represents the motif of the pattern.

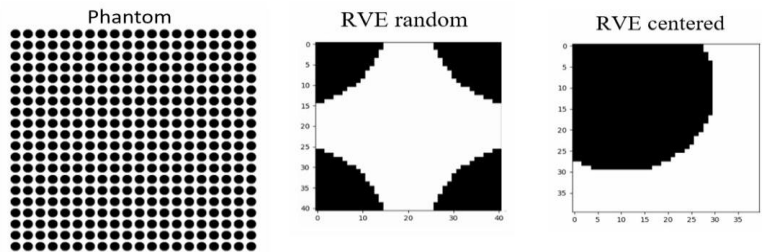

Figure 9. Highly periodic phantom and its RVE calculated with $\epsilon = 0.9$ and 10 000 rows; RVE cropped in a random place and in the position where the porosity within RVE is close to the sample porosity.

## 4. CONCLUSION

Most of the calculations conducted on Micro-CT reconstructed images are time-consuming. To speed up them the reconstructed data should be shortened, i.e. the representative volume element, the smallest informative part, should be used instead of the full domain. There was established that representative volume element and representative elementary volume are not equal definitions. In contrast to the volume, the element is cropped from a certain position in the domain that is computed according to the relevant average properties.

The robust technique of representative volume element identification in case of porous media was proposed in this paper. It enables us to find the elementary volume even in noisy, grayscale, multiphase Micro-CT reconstructed images of porous samples basing on the feature's morphology and their spatial distribution. The method is adjustable by the precision parameter that makes it possible to vary the volume size if it is needed, e.g. on denoising purposes.

The work of the algorithm was examined on diverse phantoms and their Micro-CT images. As a result, the optimal sampling frame, several voxel rows from the images that utilized as an input parameter for the algorithm, was determined and constitutes 10 000 rows. The size of the representative volume element is reducing when the precision parameter increasing and vice versa. It completely fits the theory. In noisy images the size is much smaller than in clear ones that agree with statements made in section 2.3. In highly periodic samples the technique works perfectly: it chooses the motif as elementary volume.

This work was partly supported by RFBR (grants 19-01-00790, 18-29-26019), by the Ministry of Science and Higher Education within the State assignment FSRC «Crystallography and Photonics» RAS in part of tomography modeling.